# Scheduling for Multi-Camera Surveillance in LTE Networks

Chih-Hang Wang
Department of Computer Science,
National Tsing Hua University,
Hsinchu 300, Taiwan
s100062591@m100.nthu.edu.tw

De-Nian Yang, Wen-Tsuen Chen
Institute of Information Science,
Academia Sinica,
Nankang, Taipei 115, Taiwan
dnyang@iis.sinica.edu.tw, chenwt@iis.sinica.edu.tw

*Abstract* — **Wireless surveillance in cellular networks has become increasingly important, while commercial LTE surveillance cameras are also available nowadays. Nevertheless, most scheduling algorithms in the literature are throughput, fairness, or profit-based approaches, which are not suitable for wireless surveillance. In this paper, therefore, we explore the resource allocation problem for a multi-camera surveillance system in 3GPP Long Term Evolution (LTE) uplink (UL) networks. We minimize the number of allocated resource blocks (RBs) while guaranteeing the coverage requirement for surveillance systems in LTE UL networks. Specifically, we formulate the Camera Set Resource Allocation Problem (CSRAP) and prove that the problem is NP-Hard. We then propose an Integer Linear Programming formulation for general cases to find the optimal solution. Moreover, we present a baseline algorithm and devise an approximation algorithm to solve the problem. Simulation results based on a real surveillance map and synthetic datasets manifest that the number of allocated RBs can be effectively reduced compared to the existing approach for LTE networks.**

*Keywords—Long term evolution (LTE), resource allocation, surveillance, approximation ratio, np-hard*

## I. Introduction

The need for faster and more reliable mobile services has focused significant attention on wireless broadband systems such as 3GPP Long Term Evolution (LTE). The LTE standardization aims at developing the future cellular technologies which can provide a high peak-data-rate, reduced latency, scalable bandwidths, and improved system capacity and coverage. In order to achieve this goal, Orthogonal Frequency Division Multiple Access (OFDMA) has been selected for LTE downlink (DL) radio access technology because of its robustness to multipath fading, higher spectral efficiency, and bandwidth scalability [28]. However, OFDMA has a high Peak-to-Average Power Ratio (PAPR), which is inefficient for the user equipment (UE) [29] since the power of UE is limited. Such undesirable high PAPR increases the cost of the UE and drain the battery faster. As a result of seeking an alternative to OFDMA, Single Carrier OFDMA (SC-FDMA) has been adopted as the LTE uplink (UL) multiple access scheme. SC-FDMA has a significantly lower PAPR which significantly benefits the UE in terms of transmit power efficiency since the underlying waveform is essentially single-carrier.

The resource allocation problem in LTE networks has been well explored in the previous works. In the LTE downlink (DL) system, some previous works modified the well-known Proportional Fair (PF) scheduler to perform joint scheduling in order to maximize system throughput while maintaining the fairness among users [1], [2]. Others took into account user demands for quality of service (QoS) by first satisfying higher priority users and then serving the remaining users [3], [4]. By contrast, the LTE uplink (UL) system operates under two inherent constraints: continuous allocation constraint and robust rate constraint [6] since its underlying waveform is essentially single-carrier. Continuous allocation constraint means that the user can only use adjacent resource blocks (RBs) while robust rate constraint means that the adjacent RBs allocated to one user must use the same MCS which is the most robust one among the adjacent RBs. Thus recent works have considered information related to adjacent subchannels while allocating RBs to users [5], [6], [7]. They achieved this goal by setting a window to determine the data rate of adjacent subchannels or by partitioning a channel into an ordered set. However, in the DL or UL neither the above scheduling and allocation schemes considered the demand for wireless surveillance.

In recent years, unexpected events, including major terrorist attacks and criminal events, have led to increased demands for sophisticated surveillance systems. In the past, such systems used high capacity wired or WiFi networks to transmit multimedia data. Today, however, in light of high spectral efficiency and scalable bandwidth management in LTE technologies, US police departments have been collaborating with wireless ISPs to roll out innovative and low-cost surveillance systems on 3G/4G networks [8], [9]. In addition, commercial LTE surveillance cameras and related products are also available in the markets [18], [19]. Therefore, LTE wireless surveillance has become increasingly important nowadays.

Nevertheless, new challenges also arise for efficient scheduling in LTE wireless surveillance. First, the bandwidth consumption is expected to soar to support a huge amount of video information from cameras, and minimizing the resource consumption thereby is very important. Second, when the

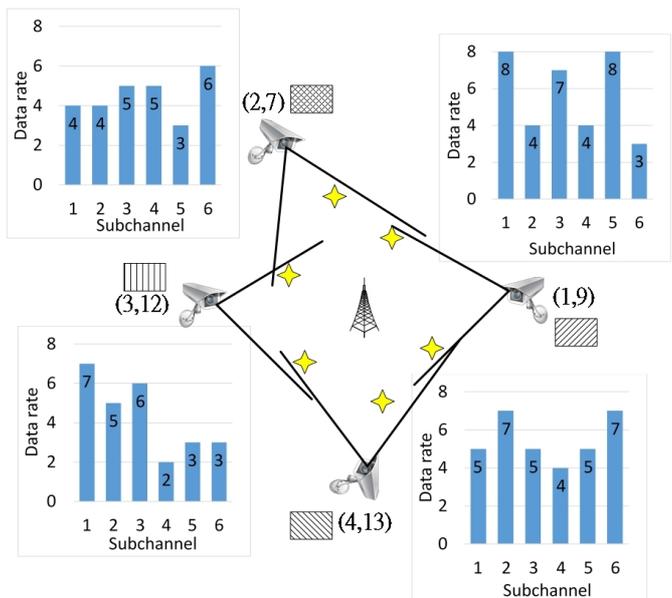

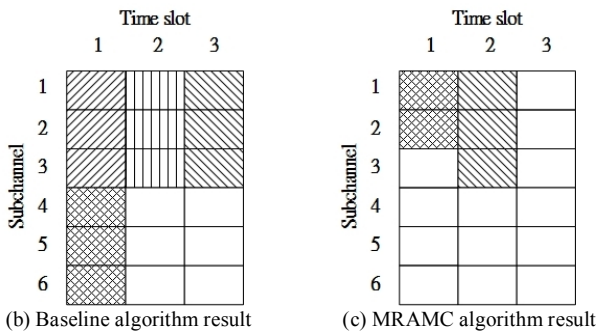

(a) Surveillance system topology and data rate information of each camera on each subchannel

(b) Baseline algorithm result

(c) MRAMC algorithm result

Fig. 1. Scheduling example for a multi-camera surveillance system.

network is congested and not able to support all cameras, the current fair scheduling strategy is not a feasible approach because the data rate shared by each camera is not be able to support the required video quality. In this situation, an efficient strategy to select suitable cameras to ensure the coverage of surveillance is desired. However, the problem is challenging because the channel condition of each uplink channel of a camera varies, while each geographic surveillance target is covered by different cameras, and each camera can observe multiple targets. Fig. 1 presents an illustrative example with four cameras, six surveillance targets, and a scheduling frame with six subchannels and three time slots. All the targets need to be covered by at least one camera and the resource allocation constraint of LTE UL should be satisfied. Fig 1(a) shows the topology and the achieved data rate of each camera on each subchannel. The traditional SNR measurement algorithm results in 12 allocated RBs as shown in fig 1(b) while the algorithm considering camera coverage results in only 6 allocated RBs as shown in fig 1(c).

In this paper, therefore, we first explore the resource allocation problem for wireless surveillance systems in heavy loaded LTE networks for a large number of surveillance cameras and then extend it to the general case afterward. More specifically, given 1) a set of wireless cameras, 2) a set of static surveillance targets (e.g., banks, post offices, etc.), 3) the channel condition of each camera in each subchannel, and 4) the number of available RBs in each time slot, we formulate a new optimization problem, named Camera Set Resource Allocation Problem (CSRAP), to minimize the number of allocated RBs for wireless surveillance cameras in a frame while guaranteeing that every geographic surveillance target can be observed by at least one camera. We prove that CSRAP is NP-Hard and inapproximable within $\ln n$, where $n$ is the number of observed surveillance targets. To solve the problem, we first study a baseline scheduling algorithm based on SNR measurements and use a small-scale example to explain why this intuitive algorithm is not suitable and why the system must explicitly consider coverage requirements. Afterward, we design a $c \ln n$ approximation algorithm, called Minimum Resource Allocation Maximum Coverage (MRAMC), where c is $\frac{r_{max}}{r_{min}}$, and $r_{max}$ and $r_{min}$ denote the best and the worst MCS rate. MRAMC iteratively selects a camera with the minimum average cost to ensure the coverage requirement and then adjusts the camera allocation to meet the scheduling constraint, which means that an RB can be allocated to only one camera. Afterward, we extend MRAMC to 1) the scenario with sufficient resources to allocate multiple cameras to the surveillance targets that are required to be observed from more angles and 2) the scenario for scheduling both the surveillance traffic and the traditional traffic. In addition to synthetic datasets, we also evaluate the above algorithms on a real surveillance camera map.

The remainder of this paper is organized as follows. Section II summarizes the related works. We introduce the system model, formulate the CSRAP, and present the hardness result in section III. Section IV first studies a baseline algorithm and then presents the proposed approximation algorithm. Section V compares different algorithms in synthetic and real datasets. Finally, we conclude this paper in section VI.

## II. RELATED WORK

### A. LTE UL Resource Allocation

Resource allocations in the LTE UL have been studied in several works [5], [6], [7]. Wong et al. [5] formulated the resource allocation problem as a set partitioning problem and maximized the weighted-sum rate for LTE UL systems. Chao et al. [6] maximized the total system throughput by considering the data rates of continuous and non-continuous scheduling. Ren et al. [7] incorporated the queue length and channel state information in a LTE UL frequency domain packet scheduling (FDPS) problem to maximize the profit function. However, the above works did not specifically consider the security requirements of multi-camera surveillance systems (e.g., coverage), and their approaches thereby tend to miss some important cameras with worse channel quality. Contrarily, our proposed algorithm can minimize the allocated RBs while guaranteeing the coverage requirement for wireless multi-camera surveillance systems in LTE networks.

## B. Video Surveillance System in Other Wireless Networks

Previous works have proposed camera systems for different applications, such as security systems, people tracking, or tele-immersive environments [10], [11], [12]. Stringa and Regazzoni [10] proposed a video-based surveillance system for detecting the presence of abandoned objects in a guarded area. Yang et al. [11] proposed a cross-layer framework with QoS-enabled streaming in a tele-immersive 3D multi-camera environment, which was later improved by considering dynamic bandwidth management in [12]. Moreover, multi-camera surveillance has been explored in WiFi and sensor networks with limited resources [13], [14], [15]. Toni et al. [13] considered the correlation between each packet and minimized the distortion in the scene reconstruction for multi-camera streaming in WiFi. Shiang and van der Schaar [14] focused on allocating the available time fraction to maximize received video quality while considering the distortion impact and delay constraints in multihop wireless networks. Durmus et al. [15] proposed an algorithm to achieve fair resource scheduling for application level messaging units in sensor networks. Wu and Hwang [16] proposed a cross-layer approach to fairly schedule all wireless cameras with varying channel quality in WiMAX, while the selection of cameras to ensure the coverage is not considered. Tseng et al. [17] minimized the number of sensors to guarantee that each object can be monitored by at least $k$ sensors satisfying some angle constraints. Nevertheless, the above approaches are not designed for LTE UL. To our best knowledge, no previous studies have examined the resource allocation problem in terms of coverage requirements for multi-camera surveillance systems in LTE networks, which has gained increasing attentions recently [8], [9].

## III. PROBLEM FORMULATION

In this section, we first introduce the system model and then formulate the Camera Set Resource Allocation Problem (CSRAP). We also propose an Integer Linear Programming formulation for CSRAP. Finally, we prove that CSRAP is NP-Hard and inapproximable within $\ln n$.

An LTE radio frame consists of time domain and frequency domain. The frame with 10ms duration is divided into ten equal-sized subframes, and each subframe is further divided into two equal-sized time slots with 0.5ms duration. A basic scheduling unit in LTE, called an RB, consists of a time slot in the time domain and a subchannel in the frequency domain, which consists of 12 consecutive subcarriers and has a 180 kHz bandwidth [20].

We consider a surveillance system that consists of $K$ cameras $\mathbb{K} = \{1,2,\dots,K\}$ and $Y$ observed static geographical objects $\mathbb{Y} = \{1,2,\dots,Y\}$. Each camera is associated with a coverage set $S_k$, which represents the set of objects covered by the camera, where $\bigcup_k S_k = \mathbb{Y}$. We define the LTE network frame duration for scheduling as $\rho$ ($\rho = 10$ms in LTE) with $T$ uplink time slots $\mathbb{T} = \{1,2,\dots,T\}$, and the scheduling algorithm runs every frame length. At each time slot $t$, a base station can allocate $M$ RBs to $K$ cameras. Specifically, a set of continuous RBs can be assigned to one camera, and each RB can be assigned to at most one camera. We denote $A_k^*$ as the collection of all sets of continuous RBs that can just achieve the rate requirement of camera $k$. In other words, for $\forall a \in A_k^*$, $a = \{i, i+1, \dots, i+l \mid mcs_a \cdot l < R_k$, and $mcs_a \cdot (l+1) \geq R_k$, $1 \leq i \leq i+l \leq M\}$, where $a$ is a set of continuous RBs, $mcs_a$ is the MCS used for allocation $a$, and $R_k$ is the requirement on the video rate for camera $k$. $\phi_{k,a,t}^\rho$ denotes the number of RBs allocated to camera $k$ by using allocation $a$ at time slot $t$ within a frame duration $\rho$. Our problem is formulated as follows:

**Problem:** The Camera Set Resource Allocation Problem
**Instance:** A set of cameras $\mathbb{K} = \{1,2,\dots,K\}$ with $K$ coverage sets $\mathbb{S} = \{S_1, S_2, \dots, S_K\}$, $M$ channel conditions for each camera, a set of objects $\mathbb{Y} = \{1,2,\dots,Y\}$ and a scheduling frame with $M \times T$ RBs in the LTE UL.
**Task:** To allocate RBs and select a set of cameras to cover all objects such that the total number of allocated RBs is minimized.

In the following, we propose an Integer Linear Programming formulation for CSRAP. The CSRAP is a minimization problem with the following objective function:

$$\text{minimize} \quad z = \sum_{k=1}^{K} \sum_{a \in A_k^*} \sum_{t=1}^{\rho} X_{k,a,t}^\rho \cdot \phi_{k,a,t}^\rho$$

and with the following constraints:

$$\sum_{k=1|y \in S_k}^{K} \sum_{a \in A_k^*} \sum_{t=1}^{\rho} X_{k,a,t}^\rho \geq 1 \quad \forall y \in \mathbb{Y} \quad (1)$$

$$\sum_{k=1}^{K} \sum_{a \in A_k^*} X_{k,a,t}^\rho \cdot \phi_{k,a,t}^\rho \leq M_t \quad \forall t \in \mathbb{T} \quad (2)$$

$$\sum_{k=1}^{K} \sum_{a \in A_k^* | m \in a} X_{k,a,t}^\rho \leq 1 \, \forall m \in \{1,\dots,M\}, \forall t \in \mathbb{T} \quad (3)$$

$$\sum_{a \in A_k^*} \sum_{t \in \mathbb{T}} X_{k,a,t}^\rho \leq 1 \quad \forall k \in \mathbb{K} \quad (4)$$

$$X_{k,a,t}^\rho \in \{0,1\} \quad \forall k \in \mathbb{K}, a \in A_k^*, \forall t \in \mathbb{T} \quad (5)$$

where $X_{k,a,t}^\rho$ (5) is a decision binary variable that is equal to 1 if camera $k$ uses allocation $a$ at time slot $t$ within a frame duration $\rho$, or 0 otherwise. Constraint (1) shows that each object is covered by at least one camera. Constraint (2) guarantees the maximum available resources at each time slot. Constraint (3) ensures that each resource block can only be scheduled to one camera and constraint (4) ensures that each camera can only use one allocation in a frame.

In the following, we prove that CSRAP is NP-Hard with the reduction from Weighted Set Cover Problem (WSCP) [23].

**Theorem 1.** *CSRAP is NP-hard*
**Proof.** Let $\mathbb{U}$ denote the set of elements. Let $\mathbb{C} = \{C_1, C_2, \dots, C_m\}$ be the collection of sets in WSCP such that each set in $\mathbb{C}$ with weight $w_i$ is a subset of $\mathbb{U}$, and $\bigcup_{i \in 1,2,\dots,m}\{C_i\} = \mathbb{U}$. The problem is to find a subset $\mathbb{C}^* \subseteq \mathbb{C}$ with $\bigcup_{C_j \in \mathbb{C}^*}\{C_j\} = \mathbb{U}$ such that the total weight $\sum_{i \in \mathbb{C}^*} w_i$ is minimized. For each set of WSCP, we construct an instance of CSRAP by defining $\mathbb{Y} = \{1,2,\dots,Y\} = \mathbb{U}$ as the set of objects to be observed, and $\mathbb{K} = \{1,2,\dots,K\} = \mathbb{C}$ as the set of cameras. Let the set of objects covered by camera $k$ as $S_k$, where $\bigcup_k S_k = \mathbb{Y}$. Each camera $k$ may require a different number of RBs, $\phi^\rho_{k,a,t}$, according to different allocation $a$ in different time slot $t$. Since CSRAP is a minimization problem, we set the smallest value of $\phi^\rho_{k,a,t}$ for camera $k$ as $w_k$.

For the WSCP, if there exists a collection of set $\mathbb{C}^*$ such that the union of elements in $\mathbb{C}^*$ is $\mathbb{U}$, we can find the corresponding subset $\mathbb{K}^* \subseteq \mathbb{K}$ such that the union of covered objects in $\mathbb{K}^*$ is also $\mathbb{Y}$. Conversely, if there exists a subset $\mathbb{K}^* \subseteq \mathbb{K}$ with the union of covered objects in $\mathbb{K}^*$ as $\mathbb{Y}$, we can find out the corresponding collection of sets $\mathbb{C}^*$ such that the union of elements in $\mathbb{C}^*$ is also $\mathbb{U}$. This is because each camera in $\mathbb{K}$ with weight $w_k$ corresponds to a covering set in $\mathbb{C}$ with weight $w_i$. Hence, the hardness of CSRAP is as hard as WSCP, and the theorem follows. ∎

From Theorem 1, CSRAP is at least as hard as the set cover problem. Feige [23] proved that the set cover problem is inapproximable within $\ln n$ and thus we have the following result.

**Corollary 1.** *CSRAP is inapproximable within* $\ln n$.

### IV. ALGORITHM DESIGN

In this section, we first describe a baseline algorithm according to SNR measurements, and explain that why this approach is not suitable for a surveillance system in the LTE UL. We then propose the Minimum Resource Allocation Maximum Coverage (MRAMC) algorithm, which minimizes the number of allocated RBs while guaranteeing the coverage requirement of the surveillance system. Afterward, we evaluate the approximation ratio of MRAMC, and analyze the space and time complexity. Finally, we extend MRAMC to 1) the scenario with sufficient resources to allocate multiple cameras to the surveillance targets that need to be observed from more angles, and 2) jointly schedule the surveillance traffic and the traditional traffic.

#### A. Baseline Scheduling Algorithm

Most scheduling algorithms adopt only the measured SNR value as the scheduling criterion. Based on this value, the scheduler selects a modulation and coding scheme (MCS), which is further mapped to the data rate for each device on each RB. The MCS determines the number of RBs required by each device. If one device chooses a higher MCS, fewer RBs is required to satisfy its data rate requirement. Hence, the baseline approach assigns the RB to the camera with the best MCS (data rate). In addition, with the continuous allocation constraint in the LTE UL, the algorithm iteratively examines the current RB and its subsequent RBs, and selects the most robust MCS for transmission. When all objects are covered, the algorithm stops.

Fig. 1 presents an example to explain why the baseline scheme is not suitable for a multi-camera surveillance system. The multi-camera surveillance system has four cameras and six surveillance targets. The scheduling frame has three time slots, and each time slot contains six RBs. The surveillance system topology and data rate requirement of each camera for each RB is shown in Fig. 1(a). The yellow star represents the surveillance target and the 2-tuple represents (*camera number, data rate requirement*). The bar chart next to the camera represents the achieved data rate of the camera on each subchannel. For the baseline scheme, the highest achieved data rate in subchannel 1 among all candidate cameras is 8. Therefore, the scheduler allocates RB1 in time slot1 to Camera1, and subsequently allocates RB2 to Camera1 since its data rate requirement is not satisfied by the allocation of RB1 alone. Afterward, the data rate for Camera1 on each allocated RB is 4, which is the most robust data rate among the allocated RBs (i.e. min(8, 4)). RB3 is also allocated to Camera1 to meet the data rate requirement. The most robust data rate among the allocated RBs is also 4 (i.e. min(8, 4, 7)), and thus the data rate requirement of Camera1 is met (i.e. 4*3 = 12 > 9). Next, because the highest achieved data rate among the remaining cameras on RB4 is 5, the scheduler allocates RB4 to Camera2. Again, the scheduler sequentially allocates RBs to Camera2 to meet its data rate requirement. The following scheduling procedure is similar to that for Camera1, which schedules the camera with the best MCS among all candidate cameras on the current RB and subsequently allocates the adjacent RB to the camera to satisfy the data rate requirement. The total number of allocated RBs for the baseline scheme is 12.

In this example, the baseline algorithm examines only the channel quality during the selection of the cameras, and the algorithm stops when all surveillance spots are covered. Hence, this algorithm tends to select more cameras and requires more RBs in LTE networks.

#### B. Proposed Scheduling Algorithm

To solve CSRAP, we propose the Minimum Resource Allocation Maximum Coverage (MRAMC) algorithm, which selects the cameras and allocates the minimum number of RBs to cover all objects. MRAMC iteratively chooses the camera with the minimum average cost to ensure that all uncovered objects can be covered with the minimum number of allocated RBs. The average cost is defined as follows:

$$\frac{\phi^\rho_{k,a,t}}{|S_k \cap \mathbb{Z}|} \tag{6}$$

where $\phi^\rho_{k,a,t}$ is the number of RBs allocated to camera $k$ with allocation $a$ in time slot $t$ of one frame size $\rho$, and $\mathbb{Z}$ are the set of objects that have not been covered. MRAMC consists of two parts: greedy scheduling and RB relocation. The first step schedules the camera without considering the scheduling constraint, which means that an RB can be reassigned to different cameras, while the second step re-allocates the RBs to meet the scheduling constraint.

*1) Greedy Scheduling*

In the greedy scheduling stage, MRAMC iteratively selects the camera with the minimal average cost. First, the scheduler finds all possible continuous RBs that can satisfy the rate requirement for each camera. The number of all possible continuous RBs for each time slot is at most $\frac{(1+M) \cdot M}{2}$ where $M$ is the number of RBs in a time slot. The scheduler then calculates the average cost for each camera over every possible continuous RBs, as defined in (6). Afterward, the scheduler

**Algorithm 1.** Minimum Resource Allocation Maximum Coverage
1: **Initialize** $\mathbb{Z} = \mathbb{U}, \mathbb{G} = \emptyset, \mathbb{K}^* = \emptyset$
**Greedy Scheduling:**
2: **repeat**
3:   Calculate $\frac{\phi_{k,a,t}^\rho}{|S_k \cap \mathbb{Z}|}$ for each $k \in \mathbb{K}, a \in A_k^*, t \in \mathbb{T}$
4:   $(k^*, a^*, t^*) = \arg min_{k \in \mathbb{K}, a \in A_k^*, t \in \mathbb{T}} \frac{\phi_{k,a,t}^\rho}{|S_k \cap \mathbb{Z}|}$
5:   $\mathbb{K} = \mathbb{K} \backslash k^*$
6:   $\mathbb{K}^* = \mathbb{K}^* \cup k^*$
7:   $\mathbb{Z} = \mathbb{Z} \backslash S_{k^*}$
8:   $\mathbb{G} = \mathbb{G} \cup (k^*, a^*, t^*)$
9: **until** $\mathbb{Z} = \emptyset$

**Adjustment Step:**
10: **repeat**
11:  Select the *unadjusted camera* $k \in \mathbb{K}^*$ with the minimum scheduling result $\phi_{k,a,t}^\rho$ from Greedy Scheduling
12:  Assign the same allocation result $a$ to $k$
13:  Set $k$ as *adjusted camera*
14:  **for** each *unadjusted camera* $k'$ whose allocation overlaps with $k$ **do**
15:   Select $min_{a' \in A_{k'}^*, t' \in \mathbb{T}} \phi_{k',a',t'}^\rho$ for each $k'$ where the RBs of $a'$ have not been occupied by any other cameras
16:   Assign $a'$ to camera $k'$
17:   set $k'$ as *adjusted camera*
18:   $\mathbb{G} = (\mathbb{G} \cup (k', a', t')) \backslash (k', a^*, t^*)$
19:  **end for**
20: **until** there is no *unadjusted camera*
21: **return** $\mathbb{G}$

selects the camera with the minimum average scheduling cost. That is, the selected camera exerts a lower average cost for an uncovered object. In this process, an RB is allowed to be assigned to more than one camera. In practical, however, an RB must be assigned to only one camera. Therefore, an adjustment step is required to relocate the cameras to meet this practical constraint.

*2) RB Relocation*

In the relocation step, we consider the scheduling constraint and relocate the RBs of some cameras, if the RBs are assigned to multiple cameras in the greedy scheduling phase. Here, an *adjusted camera* represents that the allocation of RBs for the camera has been adjusted, while other cameras whose allocated RBs overlap with the other cameras selected in the previous step are *unadjusted cameras*. The algorithm iteratively adjusts *unadjusted cameras* as follows.

1. The scheduler selects an unadjusted camera whose scheduling result from greedy scheduling phase has the minimum number of allocated RBs. The camera is assigned the same RBs.
2. Then, the scheduler iteratively relocates the RBs of unadjusted cameras, which overlaps the RBs of the unadjusted camera selected in step 1, to other RBs that have not been occupied by any cameras. The scheduler selects the RBs with the minimum number that can satisfy the rate requirement of the camera.
3. The scheduler sets the unadjusted cameras considered in the above two steps as adjusted cameras and finds the next unadjusted camera. The above process is repeated iteratively until there is no unadjusted camera.

In the following, we compare MRAMC and the baseline approach in the example in Fig. 1. MRAMC first considers RB1 in time slot1 and schedules Camera2 due to its lowest average cost (Camera1 is 3/2, Camera2 is 2/3, Camera3 is 3/2, and Camera4 is 3/3). The average costs are computed for each camera according to all possible continuous RB allocations that can just achieve the rate requirement of the camera. Then, the scheduler updates the average cost of each camera (Camera1 is 3/1, Camera3 is 3/1 and Camera4 is 3/3). Camera4 is selected and the algorithm stops since all objects are covered. The total number of allocated RBs is 5, i.e., less than half of the RBs allocated in the baseline scheme. Fig. 1(d) depicts the optimal result with the proposed algorithm. The pseudo code for MRAMC is presented in Algorithm 1.

*C. Approximation Ratio*

In this section, we derive the approximation ratio of MRAMC. We first study the approximability of MRAMC without the scheduling constraint and then derive the approximation ratio for the general case. More specifically, let $N_s$ and $N_{ws}$ denote the number of RBs for MRAMC with and without the scheduling constraint. Let $N_s^*$ and $N_{ws}^*$ represent the optimal solution with and without the scheduling constraint. Let $d_k$ denote number of objects in the coverage range of camera $k$, and $d^* = \max_k d_k$. $H(d^*)$ is the harmonic function of $d$, i.e., $H(d^*) = \sum_{i=1}^{d^*} 1/i = \ln d^* + \Theta(1)$.

**Lemma 1.** For CSRAP without the scheduling constraint, $N_{ws} \leq H(d^*) \cdot N_{ws}^*$.

**Proof.** For each camera $k$, let $S_k = \{1, 2, \dots, d_k\}$ denote the set of objects in the coverage range of camera $k$. Since each object can be covered by multiple cameras, the objects in $S_k$ are sorted according to their cover sequence in our algorithm, which means that $i < j$ if object $i$ is covered before object $j$ in MRAMC, $1 \leq i \leq d_k, 1 \leq j \leq d_k$. In other words, for the iteration that $j$ is covered, the objects in $S_k$ that have not been covered at the beginning of the iteration are $\{j, j+1, \dots, d_k\}$. Since the average cost of each element in set $S_k$ is defined in (6), we have $\frac{\phi_{k,a,t}^\rho}{|S_k \cap \mathbb{Z}|} = \frac{\phi_{k,a,t}^\rho}{d_k - j + 1}$ where $\mathbb{Z}$ is the set of objects that have not been covered at the beginning of the iteration.

Suppose $j$ is covered when set $S_{k'}$ is selected in MRAMC. The cost of $j$ is as follows.

$$cost_j = \frac{\phi_{k',a,t}^\rho}{|S_{k'} \cap \mathbb{Z}|} \leq \frac{\phi_{k,a,t}^\rho}{|S_k \cap \mathbb{Z}|} = \frac{\phi_{k,a,t}^\rho}{d_k - j + 1} \qquad (7)$$

The inequality in (7) holds because the algorithm chooses the set with the minimum average cost in each iteration. Let the total cost of the objects in set $S_k$ selected by the algorithm as $\sum_{j \in S_k} cost_j$. The upper bound of the cost for $S_k$ is then obtained as

$$\sum_{j \in S_k} cost_j = \sum_{j=1}^{d_k} cost_j \leq \sum_{j=1}^{d_k} \frac{\phi_{k,a,t}^{\rho}}{d_k - j + 1} = H(d_k) \cdot \phi_{k,a,t}^{\rho} \quad (8)$$

TABLE I. SIMULATION SETTINGS

| Parameter | Setting | Parameter | Setting |
|---|---|---|---|
| Bandwidth | 10MHz | Bandwidth per RB | 180 kHZ |
| Channel model | Typical Urban | Shadowing | Log-normal |
| Schedule length | 1 frame | TX power | 24dBm |
| Monitored area | 500*500 m² [17] | Object distribution | Uniform Cell-Edge |
| Camera Type and View distance | Omnidirectional: 30m~60m [25] Directional: 60m~100m [26] | MCS settings | QPSK、16QAM[22] |

The inequality holds due to (7), and the last equality holds since $H(d_k)$ is the harmonic function.

According to (8), we have $\phi_{k,a,t}^{\rho} \geq \frac{1}{H(d_k)} \cdot \sum_{j \in S_k} cost_j \geq \frac{1}{H(d^*)} \cdot \sum_{j \in S_k} cost_j$ where $d^* = \max_k d_k$. Let $\mathbb{G}^*$ denote the optimal solution, and $\mathbb{G}$ denotes the solution of MRAMC. Therefore, the total cost of the optimal solution follows the inequality below.

$$N_{ws}^* = \sum_{k \in \mathbb{G}^*} \phi_{k,a,t}^{\rho} \geq \frac{1}{H(d^*)} \cdot \sum_{k \in \mathbb{G}^*} \sum_{j \in S_k} cost_j \quad (9)$$

Furthermore, since each object in the universal set $\mathbb{U}$ is covered at least once by the cameras in $\mathbb{G}^*$, we have,

$$\sum_{k \in \mathbb{G}^*} \sum_{j \in S_k} cost_j \geq \sum_{j \in \mathbb{U}} cost_j = \sum_{k \in \mathbb{G}} \phi_{k,a,t}^{\rho} = N_{ws} \quad (10)$$

Finally, we obtain the upper bound for the algorithm by (9) and (10), $N_{ws} = \sum_{j \in \mathbb{U}} cost_j \leq H(d^*) \cdot N_{ws}^*$. ∎

**Theorem 2.** For CSRAP with the scheduling constraint, $N_s \leq \frac{r_{max}}{r_{min}} \cdot H(d^*) \cdot N_s^*$.

**Proof.** In the second phase of MRAMC, we relocate the RBs of some cameras selected in the first phase to meet the scheduling constraint. The worst case is that the cameras will switch from the RBs with the optimal channel conditions to those with the worst channel conditions and thus induce additional resource consumption. The number of RBs allocated to a camera is based on the MCS rate. Therefore, the adjustment for one camera will exert at most $\frac{r_{max}}{r_{min}}$ RBs, where $r_{max}$ and $r_{min}$ denote the best and the worst MCS rates. Therefore, $N_s \leq \frac{r_{max}}{r_{min}} \cdot N_{ws}$. Moreover, since the feasible region of CSRAP without the scheduling constraint is larger than the one with the scheduling constraint, $N_{ws}^* \leq N_s^*$ holds for the minimization problem CSRAP. Therefore, $N_s \leq \frac{r_{max}}{r_{min}} \cdot N_{ws} \leq \frac{r_{max}}{r_{min}} \cdot H(d^*) \cdot N_{ws}^* \leq \frac{r_{max}}{r_{min}} \cdot H(d^*) \cdot N_s^*$, and the theorem follows. ∎

### D. Complexity Analysis

In this section, we analyze the time and space complexity of MRAMC. Since MRAMC maintains the information of each camera on each RB and the observed objects in the environment, the space complexity is $O(KM + N)$.

As for time complexity, to find the camera with the lowest average cost, the scheduler first computes all possible continuous RBs that can satisfy the rate requirement for each camera of each time slot. This takes at most $K \cdot \frac{(1+M) \cdot M}{2} \cdot T = O(KM^2T)$ time for $K$ cameras. Then, the scheduler iteratively selects the camera with the smallest average cost until each object is covered by at least one camera. The algorithm takes $O(K)$ iterations to guarantee the coverage since there are at most $K$ cameras. Moreover, the minimum average cost can be found in $O(\log(KM^2T))$ time by using a priority queue in each iteration. Thus, the time complexity for greedy scheduling phase is $O(KM^2T) + O(K \log(KM^2T)) = O(K \log(KM^2T))$. In the adjustment step, the scheduler first takes $O(K)$ time to find the unadjusted camera with the smallest number of allocated RBs. Then, the scheduler requires $O(KM^2T \cdot \log M^2T))$ time to sort the allocations for $K$ cameras based on the number of RBs. Afterward, the scheduler takes $O(K)$ time for the RBs of unadjusted cameras. The time complexity of adjustment step is $O(K) + O(KM^2T \log(M^2T)) + O(K) = O(KM^2T \log(M^2T))$. As a result, the total time complexity of MRAMC is $O(K \log(KM^2T)) + O(KM^2T \log(M^2T))$.

### E. Discussion

In previous subsections, we propose MRAMC to assign at least one camera for each surveillance object. When the resources are sufficient, it is promising to allocate more cameras to the surveillance targets that are required to be observed from more angles. To address this issue, we slightly modify MRAMC and present *m*-MRAMC to choose the camera with the minimal resource consumption, instead of the lowest average cost. Initially, MRAMC is applied to select at least one camera to cover each target. At the second iteration, for each target that is currently covered by only one camera and able to be covered by more cameras, *m*-MRAMC finds a new camera with the minimal resource consumption, while every other target that has been covered by at least two cameras will not be considered in this iteration. If there are still sufficient resources, *m*-MRAMC is repeated to find a new camera with the minimal resource consumption for each target that needs three or more cameras. The above process stops when all available RBs are employed.

In the following, we extend proposed algorithm to jointly schedule the surveillance traffic and the traditional traffic in *m*-MRAMC. As mentioned above, *m*-MRAMC iteratively allocates the camera with the minimal resource consumption to the surveillance targets that are required to be observed from more angles. To jointly schedule the surveillance traffic and the traditional traffic, we define a parameter α which can be determined by network operators to decide the priority (importance) of different traffic. Algorithm *m*-MRAMC iteratively considers the traffic (surveillance or traditional traffic) with minimum $\frac{\phi_{i,a,t}^{\rho}}{\alpha}$, where $\phi_{i,a,t}^{\rho}$ is the number of RBs allocated to traffic *i* with allocation *a* at time slot *t* of one frame size $\rho$, where traffic *i* belong to either surveillance traffic or traditional traffic. With a larger α, traditional traffic *i* tends to be schedule earlier in the algorithm. Consequently, *m*-MRAMC jointly schedules both the surveillance and the traditional traffic by minimizing the number of allocated RBs in the system.

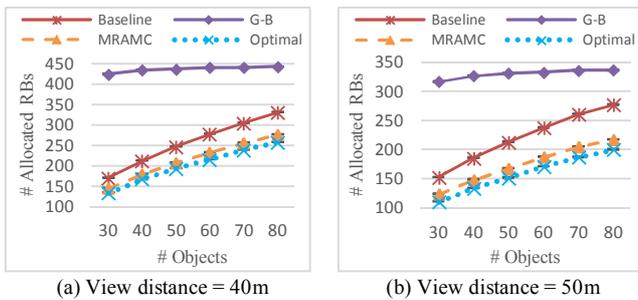

(a) View distance = 40m  (b) View distance = 50m

Fig.2. Performance of the three approaches for various numbers of objects in *overall coverage* 500*500 m² area under different view distances.

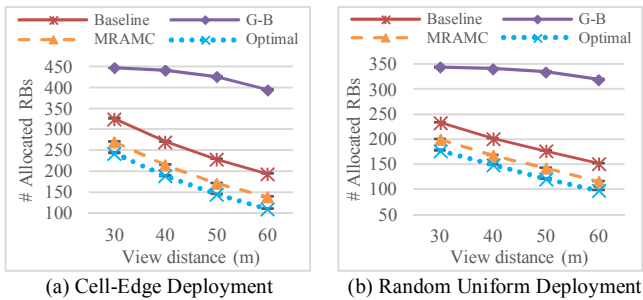

(a) Cell-Edge Deployment  (b) Random Uniform Deployment

Fig. 3. Performance of the three approaches for various omnidirectional camera view distances under different deployment schemes.

## V. SIMULATION

In this section, we compare different resource scheduling approaches for wireless surveillance in LTE networks.

### A. Simulation Setups

We consider a guarded area of $500 \times 500\ m^2$ with a base station at the center, and randomly distribute $N$ objects and $M$ surveillance cameras over the area. We consider two coverage cases: *overall coverage* and *partial coverage*. For *overall coverage*, we deploy cameras in a grid-based scheme such that the cameras can cover the entire area. For *partial coverage*, to guarantee that all objects can be covered by at least one camera, we first randomly distribute the objects and then randomly deploy a camera to cover each object. Afterward, the other cameras are randomly distributed over the area. In addition, we exploit a real surveillance map of University of Maryland [26]. In the map, we assume that each important place (e.g., intersections, buildings, or parking lots) is a surveillance spot.

The bandwidth of the LTE network is set at 10MHz with 50 RBs in each time slot. The uplink bandwidth per RB is 180 kHz, which is equal to the bandwidth of an OFDMA RB in LTE [20]. The transmission power is set at 24dbm. For a realistic simulation, the path loss, shadowing model, and MCS are based on 3GPP specifications [21], [22]. We consider omnidirectional [24] and directional cameras [25], and assume that the cameras are all capable of LTE interface. The view distance of an omnidirectional camera is set at 30m~60m, while that of a directional camera is set at 60m~100m. The simulation settings are summarized in Table 1.

In the simulation, we compare the baseline and MRAMC algorithms with the Greedy-Based (G-B) algorithm [7], where the G-B algorithm stops after all objects are covered. In addition, we also find the optimal solution with the proposed Integer Linear Programming formulation solved by Gurobi [27]. To evaluate our proposed algorithms, we change the following parameters: 1) number of objects, 2) view distance, 3) field of view (FOV), and 4) deployment scenario. We evaluate different algorithms to find the number of RBs transmitted by the base station. Each result is averaged with 5000 times.

### B. Simulation Results

#### 1) Omnidirectional Cameras

We first consider *overall coverage* and investigate the number of objects versus the number of allocated RBs for the three approaches given different camera view distances (40m and 50m). In Fig. 2, increasing the number of objects generally raises the number of allocated RBs, that is, more cameras need to be selected to cover all objects. Also, the performance gap between MRAMC and the baseline scheme increases as the number of objects grows because MRAMC minimizes the average number of RBs required to cover an object. However, it has a smaller impact on the G-B scheme since this scheme only considers channel quality in algorithm design. In Fig. 2(b), we increase the view distance, finding a trend similar to that in Fig. 2(a) except that the performance gap between MRAMC and baseline scheme grows. This indicates that, MRAMC benefits more from the cameras with greater coverage, because it minimizes the average number of allocated RBs required to cover an object.

Next, we consider *partial coverage*. In Fig. 3, we investigate the view distance of omnidirectional cameras versus the number of allocated RBs under two possible deployment scenarios [1], namely *Cell-edge* and *Random*. We set the number of cameras as 50, and the number of objects as 40. In the two deployment scenarios, MRAMC outperforms both the baseline and G-B schemes. When the view distance (i.e., camera coverage) is increased, fewer RBs are needed. In *Cell-edge* deployment, more RBs are allocated to the cameras due to the poor channel conditions, as shown in Fig. 3(a). In the *Random* deployment, the objects and cameras are equally distributed over the whole cell area, and thus the cameras can cover the objects in any direction within their view distance. Therefore, fewer cameras are involved to meet the coverage requirement for surveillance systems, as depicted in Fig. 3(b).

#### 2) Directional Cameras

Directional cameras usually have a higher resolution than omnidirectional cameras, providing a clear view at a long distance (e.g., the camera described in [24] features a night view distance of up to 100m). Fig. 4 presents the view distance of directional cameras versus the number of RBs used for the three approaches under the two deployment scenarios described in the previous subsection. The view distance ranges between 60m~100m, and the default FOV (i.e., the angle that a camera can monitor in a single frame) value is set at $120°$. In the two deployment scenarios, MRAMC outperforms both the baseline and G-B schemes. When the view distance is increased, fewer

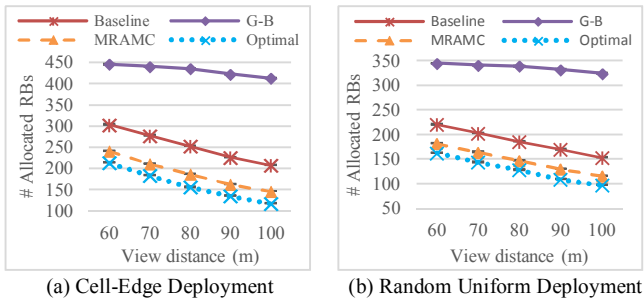

(a) Cell-Edge Deployment    (b) Random Uniform Deployment

Fig. 4. Performance of the three approaches for various directional camera view distances under different deployment schemes.

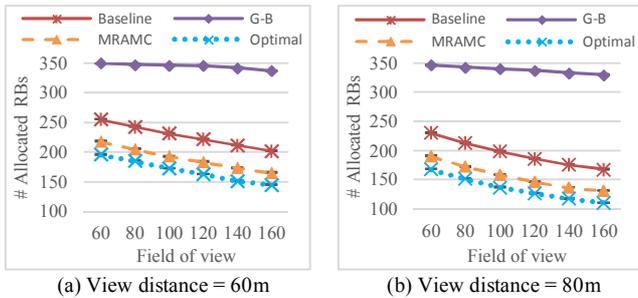

(a) View distance = 60m    (b) View distance = 80m

Fig. 5. Performance of the three approaches for various FOV values of directional cameras at different view distances.

RBs are allocated to ensure coverage. From Fig. 3 and Fig. 4, when the view distance is the same, the results manifest that directional cameras need more RBs in both deployment scenarios since omnidirectional cameras have better coverage. Fig. 5 shows the impact of FOV on the number of allocated RBs for different camera view distances. The cameras and objects are distributed randomly over the area. The number of allocated RBs is highly dependent on the camera FOV. When FOV is increased (i.e., when a camera has a wider view range), fewer cameras are needed to ensure the coverage requirement. The curves for the baseline scheme and MRAMC decrease rapidly. By contrast, the improvement of G-B is much minor since only the channel condition is examined.

*3) Real Surveillance Map*

In the following, we present the results of a real surveillance map of University of Maryland [26]. We assume that the view distance of cameras is 100m, and the FOV varies between 100~300m. Fig. 6 presents the impacts of the view distance and FOV on the number of allocated RBs. In Fig. 6(a), initially the curves decline quickly and become saturated after exceeding 200 degrees. Since the cameras are deployed near the objects, at angles exceeding 200 degrees, further increasing the FOV will not significantly add to the number of objects included. In Fig. 6(b), we set the FOV to 300 degrees, with a view distance ranging between 80m ~120m in increments of 20m. The curves decline quickly after the view distance exceeds 100m because a camera can cover a distant objects, thus fewer cameras are actually needed to cover all objects. In general, both MRAMC and baseline outperform the G-B scheme about 30%~40% and 25%~35% respectively. Therefore, the result shows that the coverage requirement must be carefully considered for the resource allocation in a surveillance system.

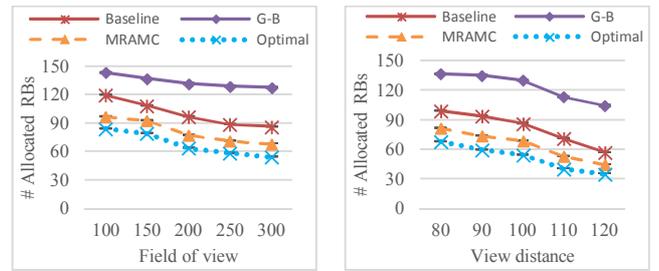

(a) View distance = 100m    (b) FOV = 300

Fig. 6. Simulation results for a real surveillance map.

## VI. CONCLUSION

Previous works on LTE resource allocation usually considered traditional network traffic. However, for surveillance system, the camera coverage should be cautiously considered such that every important spot is under surveillance. In this paper, a multi-camera surveillance system in the LTE UL is proposed to minimize the number of RBs allocated to cameras while guaranteeing surveillance system coverage requirements. We formulate the Camera Set Resource Allocation Problem (CSRAP) and prove that the problem is NP-Hard and not approximable within $ln\ n$. To solve the problem, we study a baseline scheduling algorithm based on SNR measurements and propose an approximation algorithm. Simulation results demonstrate that the number of allocated RBs can be effectively reduced compared to the existing approach for LTE networks.